# Perpendicularly Polarized Spin Hall Effects Induced by Spin-Dependent Scattering in Ferromagnetic Metals

Yuta Yahagi[1]*, Daisuke Miura[1], and Akimasa Sakuma[1]

[1]*Department of Applied Physics, Tohoku University, Sendai 980-8579, Japan*

Spin currents in ferromagnets afford diverse functionalities. We evaluate the extrinsic spin Hall effects of magnetic impurity scattering in ferromagnetic metals. We show that spin-dependent scattering can provide a high spin current polarized perpendicularly to the magnetization direction and is a dominant mechanism in the moderate-conductivity regime. We find that the superposition of the spin-conserve and spin-flip channels causes the spin currents. These findings suggest that optimizing alloy composition is an effective strategy to control the spin Hall effect.

The spin Hall effect (SHE) is a phenomenon in which a charge current is converted to a spin current and is considered a promising method for exerting spin-orbit torque (SOT) for high-efficiency magnetization switching [1–4]. It has been widely studied in nonmagnetic metals (NMs), such as platinum. Recently, the SHE in ferromagnetic metals (FMs) has attracted considerable attention because these materials exhibit higher conversion efficiency [5–12] and richer physics compared to those exhibited by NMs [13–20].

In NMs, the spin current must be polarized perpendicular to the flow direction, restricting the device geometry. In contrast, FMs can generate spin currents polarized in arbitrary directions depending on magnetization [21–23], thus facilitating the field-free SOT switching of a perpendicularly magnetized film [24–27] or the self-induced SOT on a single-layer FM film [28–32]. These unique features improve device applications by simplifying structures or even implementing new functionalities.

To express the SHE, we define the spin Hall conductivity (SHC) tensor $\sigma_{ij}^{\mu}$ as $J_i^{\mu} = \sigma_{ij}^{\mu} E_j$ with the applied electric field $E_j$ and observe the spin current $J_i^{\mu}$ flowing in an $i$-direction polarized $\mu$-direction. Let $\boldsymbol{\sigma}_{ij} = \sigma_{ij}^{\parallel}\hat{\boldsymbol{m}} + \boldsymbol{\sigma}_{ij}^{\perp} \times \hat{\boldsymbol{m}}$ be the SHC tensor in the vector form with respect to spin polarization, where $\sigma_{ij}^{\parallel}$ ($\boldsymbol{\sigma}_{ij}^{\perp}$) denotes the parallel (perpendicular) SHC whose spin polarization is parallel (perpendicular) to the magnetization direction $\hat{\boldsymbol{m}}$. From a microscopic viewpoint, $\sigma_{ij}^{\parallel}$ can be expressed as the spin-polarized current associated with the anomalous Hall effect (AHE) or the anisotropic magnetoresistance (AMR) effect [33]. By contrast, $\boldsymbol{\sigma}_{ij}^{\perp}$ cannot be expressed by a simple analogy with the AHE or AMR because it carries a perpendicularly polarized spin that is zero in the absence of an external field. Despite the counterintuitive feature of $\boldsymbol{\sigma}_{ij}^{\perp}$, considerable effects have been reported from *ab initio* calculations [34–39]. These studies suggest the importance of $\boldsymbol{\sigma}_{ij}^{\perp}$ in FM.

However, these *ab initio* studies assumed either a clean-limit or spin-independent scattering, and the impacts of spin-dependent scattering have not been fully understood. Moreover, it is controversial whether $\boldsymbol{\sigma}_{ij}^{\perp}$ can appear in actual FMs with short spin dephasing lengths. Although Amin and Davidson *et al.* provided an answer to this question within a part of the intrinsic mechanisms [22,36], the role of impurity scattering remains unclear. Further investigations on spin-dependent scattering mechanisms are required.

Some theoretical studies have explored the contributions of spin-dependent (s-d) scattering [40,41]; however, they targeted either a nonmagnetic or weak exchange splitting system and assumed a phenomenologically induced scattering potential. However, these assumptions are not valid for typical, strongly splitting FMs such as Fe,

Co, Ni, and their alloys. Furthermore, not much is known about the perpendicular SHE in typical FMs that exhibit realistic scattering processes, such as s-d scattering.

This study evaluated the SHC owing to spin-dependent scattering mechanisms based on electron theory. We consider a polycrystalline FM system in a strong exchange splitting regime and with s-d scattering, interpreted by the impurity Anderson Hamiltonian. To formulate the SHC, we employed a microscopic linear-response theory based on the Kubo formula [42]. We previously investigated only the coherent term of $\boldsymbol{\sigma}_{ij}^{\perp}$ within the same framework [43]; however, in this study, we include all the ignored terms and perform a quantitative analysis.

Throughout this study, we represent the SHC tensor by the rotational frame $\{X, Y, Z\}$ such that $\hat{\boldsymbol{m}} \parallel \hat{\boldsymbol{Z}}$. Hence, $\sigma_{ij}^Z$ corresponds to $\sigma_{ij}^{\parallel}$, and $(\sigma_{ij}^X, \sigma_{ij}^Y)$ corresponds to $\boldsymbol{\sigma}_{ij}^{\perp}$. In a polycrystalline FM, the $\hat{\boldsymbol{m}}$ dependence is entirely contained in the rotational matrix $R = R(\hat{\boldsymbol{m}})$ such that $\sigma_{nm}^l = \sum_{ijk} R_{ni} R_{mj} R_{lk} \sigma_{ij}^k$, where $\sigma_{ij}^k$ and $\sigma_{nm}^l$ denote the SHC tensors in the rotational and laboratory frames, respectively.

The Hamiltonian is given by $H = H_s + H_d + H_{\text{hyb}} + U_{\text{scat}}$, where $H_s = \sum_{\boldsymbol{k}} \sum_{\sigma} \varepsilon_{\boldsymbol{k}\sigma} c_{\boldsymbol{k}\sigma}^{\dagger} c_{\boldsymbol{k}\sigma}$ is the conduction state, represented as a plane wave with $\varepsilon_{\boldsymbol{k}\sigma} = \hbar^2 k^2 / 2m_e - \chi_\sigma \Delta_s / 2$. $c_{\boldsymbol{k}\sigma}^{\dagger}$ ($c_{\boldsymbol{k}\sigma}$) is the creation (annihilation) operator of a conduction state with momentum $\boldsymbol{k}$ and spin $\sigma = \uparrow, \downarrow$, $\hbar^2 k^2 / 2m_e$ is the kinetic energy, $\Delta_s$ is the exchange splitting, and $\chi_\sigma = \pm 1$ with $+(-)$ for $\sigma = \uparrow (\downarrow)$.

$H_d = \sum_{m,m'} \sum_{\sigma,\sigma'} (\tilde{E}_\sigma^d \delta_{\sigma',\sigma} \delta_{m',m} + V_{m'\sigma';m\sigma}^{\text{so}}) d_{m'\sigma'}^{\dagger} d_{m\sigma}$ is the localized states represented as an impurity Anderson model in a scheme of the Hartree–Fock approximation [44]. $d_{m\sigma}^{\dagger}$ ($d_{m\sigma}$) is the creation (annihilation) operator of a localized d-state with magnetic quantum number $m = -2, \ldots, 2$. The energy level is given by $\tilde{E}_\sigma^d \equiv E^d - \chi_\sigma \Delta_d / 2$ with the original energy level $E^d$ and exchange splitting $\Delta_d$. We ignore the crystal-field splitting among the d-orbitals in this study. $V_{m'\sigma';m\sigma}^{\text{so}} = \xi_{\text{so}} \{m\sigma \delta_{\sigma',\sigma} \delta_{m',m} + \sqrt{(2 - \chi_\sigma m)(3 + \chi_\sigma m)} \delta_{\sigma',-\sigma} \delta_{m',m+\sigma}\}$ is the intra-atomic SOI in the d-states. Here, the coupling constant $\xi_{\text{so}}$ is determined using the impurity element. Because $\xi_{\text{so}}$ is typically small in d-band metals, we treat $\hat{V}^{\text{SO}}$ perturbatively.

$H_{\text{sd}} = \Omega^{-1/2} \sum_i \sum_{\boldsymbol{k}} \sum_m \sum_\sigma (e^{-i\boldsymbol{k} \cdot \boldsymbol{R}_i} V_{\boldsymbol{k}m\sigma}^{\text{sd}} c_{\boldsymbol{k}\sigma}^{\dagger} d_{m\sigma} + \text{h.c.})$ is the s-d hybridization Hamiltonian, where $i = 1, \ldots, N_{\text{imp}}$ is the site index of the impurity with position $\boldsymbol{R}_i$. $N_{\text{imp}}$ is the number of impurities. $\Omega$ is the volume of the system. In this study, we assume $n_{\text{imp}} \equiv N_{\text{imp}}/\Omega \ll 1$. $V_{\boldsymbol{k}m\sigma}^{\text{sd}} = \sqrt{4\pi} V_{k\sigma} Y_{2m}(\hat{\boldsymbol{k}})$ is the s-d hybridization coefficient, where $Y_{lm}(\hat{\boldsymbol{x}})$ represents the spherical harmonics, and $V_{k\sigma}$ depends only on $k = |\boldsymbol{k}|$.

$U_{\text{scat}} = \Omega^{-1}\sum_i\sum_{k,k'}\sum_\sigma e^{-i(k-k')\cdot R_i}\left(u_{\text{imp}} + J^{\text{p}}_{k,k'}\right)c^\dagger_{k\sigma}c_{k'\sigma}$ is the spin-independent scattering potential. $u_{\text{imp}}$ is the s-s scattering potential, assumed to be constant. $J^{\text{p}}_{kk'} = 4\pi I_{\text{sp}}\sum^1_{p=-1}Y_{1p}(\hat{k})Y^*_{1p}(\hat{k}')$ denotes the scattering between a conduction state and a localized p-orbital partial wave with the coupling constant $I_{\text{sp}}$. This term is typically small; however, it can provide a leading-order contribution as the current vertex correction (CVC) term [45,46].

The velocity operator is given by $\mathcal{V}_{k\sigma} = \mathcal{V}^{\text{ss}}_{k\sigma} + \left(\sqrt{4\pi/\Omega}\sum_i e^{ik\cdot r_i}\sum_m \mathcal{V}^{\text{sd}}_{km\sigma} + \text{h.c.}\right)$, where $\mathcal{V}^{\text{ss}}_{k\sigma} = (\hbar k/m_e)\hat{k}c^\dagger_{k\sigma}c_{k\sigma}$, and $\mathcal{V}^{\text{sd}}_{km\sigma} = (V_{k\sigma}/\hbar k)\{\beta_{k\sigma}\hat{k}Y_{2m}(\hat{k}) + \Psi_{2m}(\hat{k})\}d^\dagger_{m\sigma}c_{k\sigma}$ with $\beta_{k\sigma} \equiv kV_k^{-1}(\partial V_k/\partial k)$, a dimensionless parameter for the momentum derivative of $V_k$, and $\Psi_{lm}(\hat{k}) \equiv k\nabla_k Y_{lm}(\hat{k})$, a momentum gradient of the spherical harmonics. $\mathcal{V}^{\text{sd}}_k$ is the anomalous velocity. Thus, we define the spin velocity as $\mathcal{V}^\mu_k \equiv \{\hat{\sigma}^\mu, \mathcal{V}\}/2$.

The SHC is represented based on the Kubo–Streda formula [47] at $T = 0$, written as $\sigma^{\mu,(\text{sd})}_{ij} \simeq (\hbar^2|e|/4\pi\Omega)\text{Tr}\langle\mathcal{V}^\mu_i G^+\mathcal{V}^0_j G^-\rangle$, where (sd) is added to identify the s-d scattering contribution, $G^\pm \equiv G(E_F \pm i0)$ is a retarded (advanced) Green's function of the total system at the Fermi energy $E_F$, and $\langle\cdots\rangle$ denotes the configurational average with respect to the impurity distribution. Here, we neglected some terms including $G^+G^+$ or $G^-G^-$ and the so-called Fermi sea term because their contribution is small in a metallic system [48–51].

Taking $O(n_{\text{imp}})$ terms under the averaged T-matrix approximation [43], we obtain the formal expressions of $\sigma^\mu_{ij}$, as shown in Fig. 1, where $G^{\text{s}\pm}_{k\sigma}$ and $G^{\text{d}\pm}_{m\sigma;m'\sigma'}$ are the clothed Green's functions of the conduction and localized states, given by $G^{\text{s}\pm}_{k\sigma} = (E_F - \varepsilon_{k\sigma} \pm i\hbar/\tau^s_\sigma)^{-1}$ and $G^{\text{d}\pm} = \left\{\left(\tilde{G}^{\text{d}\pm}\right)^{-1} - V^{\text{so}}\right\}^{-1}$, respectively, with $\tilde{G}^{\text{d}\pm}_{m\sigma;m'\sigma'} \equiv \delta_{m',m}\delta_{\sigma',\sigma}\left(E_F - \tilde{E}^{\text{d}}_\sigma - \Gamma^{\text{d}\pm}_\sigma\right)^{-1}$, the unperturbed term with respect to the SOI. Because this is independent of $m$ in this model, we omit $m$ from the subscript. $1/\tau^s_\sigma = 1/\tau_0 + 1/\tau^{\text{sd}}_\sigma$ is the conduction electron relaxation time. $1/\tau_0$ is that from s-s and s-p scattering, and $1/\tau^{\text{sd}}_\sigma = -n_{\text{imp}}V^2_{k\sigma}\text{Im}\tilde{G}^{\text{d}+}_\sigma$ is that from s-d scattering. $\Gamma^{\text{d}\pm}_\sigma \equiv n_{\text{imp}}(2\pi^2)^{-1}\int dk\, k^2 V^2_{k\sigma}G^{\text{s}\pm}_{k\sigma}$ is the self-energy of the localized electrons owing to s-d hybridization. In the derivation, we used the orthogonality of the spherical harmonics, for instance, $V^{\text{ds}}_{mk\sigma}G^{\text{s}}_{k\sigma}J^{\text{p}}_{k,k'} \propto \int d\hat{k}\, Y^*_{2m}(\hat{k})Y_{1p}(\hat{k}) = 0$, and also the selection rule of the CVC in which the momentum integral becomes nonzero only for a vertex among different parity [45,46], such as $V^{ds}_{mk'\sigma}G^{s}_{k'\sigma}\mathcal{V}^{ss}_{k'\sigma}G^{s}_{k'\sigma}J^{p}_{k'k} \propto \int d\hat{k}' Y^{*}_{2m}(\hat{k}')Y_{1p}(\hat{k}')\hat{k}' \neq 0$. Additionally, we considered the lowest-order terms of $I_{sp}$ as it is sufficiently small.

Figure 1 shows the leading-order terms of $\boldsymbol{\sigma}^{\perp,(sd)}_{ij} = \left(\sigma^{X,(sd)}_{ij}, \sigma^{Y,(sd)}_{ij}, 0\right)$. We omitted the terms having $\mathcal{V}^{\mu,ss}_i G \mathcal{V}^{0,sd}_j G$ or $\mathcal{V}^{\mu,sd}_i G \mathcal{V}^{0,sd}_j G$ because they are proportional to $1/\tau^{s}_{\sigma}$, which is typically negligible in a metallic system. We now discuss the microscopic origins of each term. The coherent term (a) corresponds to the anisotropic spin-flip scattering mechanism, which is a perpendicularly polarized version of the AMR-driven spin current, as described in our previous work [43]. Similarly, other terms can be understood as perpendicularly polarized versions of (c) skew-scattering with spin-flip and (b, d) side-jump with spin-flip, which have been conventionally regarded as origins of $\sigma^{\parallel}_{ij}$.

We now derive the analytical expressions for each in Fig. 1. According to a symmetry analysis, the general SHC tensor for $\sigma^{\perp}_{ij}$ is reduced to four independent components in a polycrystalline FM [43], which can be written as:

$$\sigma^X = \begin{pmatrix} 0 & 0 & \sigma_{o,+} \\ 0 & 0 & \sigma_{e,+} \\ \sigma_{o,-} & \sigma_{e,-} & 0 \end{pmatrix}, \sigma^Y = \begin{pmatrix} 0 & 0 & -\sigma_{e,+} \\ 0 & 0 & \sigma_{o,+} \\ -\sigma_{e,-} & \sigma_{o,-} & 0 \end{pmatrix}, \quad (1)$$

with $\sigma_{o,\pm} \equiv \sigma_{o,s} \pm \sigma_{o,a}$ and $\sigma_{e,\pm} \equiv \sigma_{e,s} \pm \sigma_{e,a}$. The subscript "e" or "o" denotes whether *even* or *odd* under the time-reversal operation, and "s" or "a" denotes whether *symmetric* or *anti-symmetric* under $\sigma^{\mu}_{ij} \to \sigma^{\mu}_{ji}$. To conduct the calculation, we expand $G^{d}_{m\sigma;m'\sigma'}$ with respect to $\xi_{so}$ up to the second-order and take terms of $O(\xi^2_{so})$ or $O(\xi_{so}I_{sp})$. Eventually, we obtain the following expressions:

$$\sigma^{(sd)}_{e,a} \simeq \frac{G_s}{4}\sum_{\sigma} \xi_{so} k^F_\sigma D^d_{\bar{\sigma}} \frac{\tau^s_\sigma}{\tau^{sd}_\sigma}\left\{\left(A^{(c)}_\sigma + A^{(d)}_\sigma\right)f_r + A^{(b)}f_i\right\}, \quad (2)$$

$$\sigma^{(sd)}_{o,a} \simeq \frac{G_s}{4}\sum_{\sigma} \chi_\sigma \xi_{so} k^F_\sigma D^d_{\bar{\sigma}} \frac{\tau^s_\sigma}{\tau^{sd}_\sigma}\left\{\left(A^{(c)}_\sigma + A^{(d)}_\sigma\right)f_i + A^{(b)}f_r\right\}, \quad (3)$$

$$\sigma^{(sd)}_{e,s} \simeq -\frac{G_s}{4}\sum_{\sigma} \chi_\sigma \xi^2_{so} k^F_\sigma D^d_{\bar{\sigma}} \frac{\tau^s_\sigma}{\tau^{sd}_\sigma}\left\{\left(B^{(a)}_\sigma f_r + B^{(b)}_\sigma f_i\right)(D^d_\uparrow - D^d_\downarrow)\pi \right. \\ \left. + \left(B^{(a)}_\sigma f_i + B^{(b)}_\sigma f_r\right)(\Lambda^d_\uparrow - \Lambda^d_\downarrow)\pi\right\}, \quad (4)$$

and

$$\sigma^{(sd)}_{o,s} \simeq \frac{G_s}{4}\sum_{\sigma} \xi^2_{so} k^F_\sigma D^d_{\bar{\sigma}} \frac{\tau^s_\sigma}{\tau^{sd}_\sigma}\left\{\left(B^{(a)}_\sigma f_i + B^{(b)}_\sigma f_r\right)(D^d_\uparrow - D^d_\downarrow)\pi \right. \\ \left. + \left(B^{(a)}_\sigma f_r + B^{(b)}_\sigma f_i\right)(\Lambda^d_\uparrow - \Lambda^d_\downarrow)\pi\right\}, \quad (5)$$

where $\bar{\sigma} = -\sigma$, $G_s \equiv |e|/2\pi$, $\hbar k_\sigma^F \equiv \sqrt{2m_e(E_F + \chi_\sigma \Delta_s)}$, $D_\sigma^d \equiv -\pi^{-1}\mathrm{Im}\tilde{G}_\sigma^{d+}$, $\Lambda_\sigma^d \equiv \pi^{-1}\mathrm{Re}\tilde{G}_\sigma^{d+}$, $f_r \equiv \cot\delta_{2\uparrow}\cot\delta_{2\downarrow} - 1$, and $f_i \equiv \cot\delta_{2\uparrow} + \cot\delta_{2\downarrow}$. $\delta_{2\sigma}$ is the spin-dependent phase shift of the s-d scattering given by $\tan\delta_{2\sigma} \equiv \mathrm{Im}\tilde{G}_\sigma^{d+}/\mathrm{Re}\tilde{G}_\sigma^{d+}$. It can be estimated using Friedel's sum rule for each spin state as $\delta_{2\sigma} = (n_\sigma^{\mathrm{imp}} - n_\sigma^{\mathrm{host}})\pi/5$, where $n_\sigma^{\mathrm{imp}}(n_\sigma^{\mathrm{host}})$ is the d-electron occupancy of the impurity (host) atom [52]. $A_\sigma^{(x)}$ and $B_\sigma^{(x)}$ are coefficients given in Table I, where $x = a, b, c, d$ corresponds to each term in Fig. 1, $\varepsilon_\sigma^F \equiv (\hbar k_\sigma^F)^2/2m_e$, $\zeta_\sigma^F \equiv (V_{k\bar{\sigma}}/V_{k\sigma})|_{k=k_\sigma^F}$, $\beta_\sigma^F \equiv \beta_{k\sigma}|_{k=k_\sigma^F}$, and $\tan\delta_{1\sigma} \equiv \pi I_{\mathrm{sp}} D_\sigma^s$ with $D_\sigma^s = m_e k_\sigma^F/2\pi^2\hbar^2$. In the derivation, we use the strong exchange coupling condition $\hbar/\tau_\sigma^s \ll \Delta_s$. The set of Eqs. (2)–(5) is the central result of this study, in which $\sigma_{ij}^{\perp,(\mathrm{sd})}$ is expressed solely by the electronic structure, except for $\tau_\sigma^s/\tau_\sigma^{\mathrm{sd}}$, which depends on the environment.

First, we discuss the dominant factors. As in Eqs. (2)–(5), $\sigma_{ij}^{\perp,(\mathrm{sd})}$ has the following factors: (1) $\xi_{\mathrm{so}}$, the intra-atomic SOC in the impurity; (2) $\tau_\sigma^s/\tau_\sigma^{\mathrm{sd}}$, the relaxation time factor; (3) $k_\sigma^F D_{\bar{\sigma}}^d \propto D_\sigma^s D_{\bar{\sigma}}^d$, which governs the probability of s-d scattering among different spin states; (4) $f_r$ and $f_i$, the factors representing the resonance structure of the spin-flip s-d scattering; (5) $A_\sigma^{(x)}$ or $B_\sigma^{(x)}$, the coefficient corresponding to the specific scattering process in Fig. 1; and (6) $D_\uparrow^d - D_\downarrow^d$ or $\Lambda_\uparrow^d - \Lambda_\downarrow^d$, the spin polarization among localized states. In short, $\sigma_{ij}^{\perp,(\mathrm{sd})}$ tends to be small in a half-metal as $D_\sigma^s D_{\bar{\sigma}}^d \sim 0$ and is high in a non-half-metallic FM.

An effective way to maximize $\sigma_{ij}^{\perp,(\mathrm{sd})}$ is to tailor the resonant scattering by choosing a combination of host and impurity elements, such that $f_r$ or $f_i$ is high. A similar strategy has already been applied to SHEs in NMs [53]; nevertheless, we expect FMs contain a greater variety of candidate materials than NMs due to their nonmonotonic dependence on the magnetic structure. It is worth exploring the different compositions of FMs.

Next, we proceed to estimate the order of $\sigma_{ij}^\perp$ for Fe-doped Ni as an example of a typical FM. For simplicity, we now consider $V_{k\sigma} \propto k^2$, which corresponds to employing a transition-metal pseudopotential [54]; then, we have $\beta_\sigma^F \simeq 2$ and $\zeta_\sigma^F \simeq 1$. Typically, in d-band FMs, $\tan\delta_{1\sigma} \simeq -0.1$ [55]. Using this, we estimate $A_\sigma^{(x)}$ and $B_\sigma^{(x)}$ as in the two rightmost columns in Table I. We choose $k_\sigma^F \sim 1.4$ Å$^{-1}$ [54] and $\xi_{\mathrm{so}} \sim 0.054$ eV [56]. We roughly set to $\Delta_s \sim 1$ eV, $E_F \sim 9.52$ eV, $D_\uparrow^d \sim 0$ eV$^{-1}$, $D_\downarrow^d \sim 0.3$ eV$^{-1}$, $f_r \sim 10$, and $f_i \sim -38$ from Fig 2.12 in Ref. [57]. To extract $D_\sigma^d$ as the localized d-state, we subtract $D_\sigma^s \sim 0.16$ eV$^{-1}$ from the calculated value. Assuming $1/\tau_0 \ll 1/\tau_\sigma^{\mathrm{sd}}$, i.e., $\tau_\sigma^s/\tau_\sigma^{\mathrm{sd}} \simeq 1$, we obtain $\sigma_{e,a}^{(sd)} \sim 2.15 \times 10^3, \sigma_{o,a}^{(sd)} \sim 4.48 \times 10^3, \sigma_{e,s}^{(sd)} \sim 1.28 \times 10^3$, and $\sigma_{o,s}^{(sd)} \simeq 3.23 \times 10^3$ in units of $\hbar/(2|e|\Omega\text{cm})$, which is comparable to the typical values of $\sigma_{ij}^{\parallel} \sim O(10^3) \, \hbar/(2|e|\Omega\text{cm})$. In Fig. 2, we show a comparison with the Ni host-lattice contribution $\sigma_{ij}^{\mu,(\text{host})}$ due to both the mechanisms of intrinsic and spin-independent scattering. Here, we use the values of pure fcc-Ni computed using the method given in Ref. [34] as $\sigma_{e,+}^{(\text{host})} \simeq 4600 \, \hbar/(2|e|\Omega\text{cm})$ and $\sigma_{o,+}^{(\text{host})}/\sigma_{ZZ} \simeq 0.003 \, \hbar/(2|e|)$, where $\sigma_{ZZ}$ is the longitudinal conductivity. For comparison, we similarly estimate $\sigma^{\parallel}$. Although this is an approximate estimation, $\boldsymbol{\sigma^{\perp,(\text{sd})}}$ can appear with a significant order of magnitude comparable with the host-lattice contribution as well as $\sigma^{\parallel}$.

Another essential point to note is the scaling relationship with $\sigma_{ZZ}$. $\sigma_{e,+}$ is independent of $\sigma_{ZZ}$, whereas $\sigma_{o,+}$ has a linear term of $\sigma_{ZZ}$ owing to $\sigma_{o,+}^{(\text{host})}$ (particularly the spin-independent mechanism [34]). Furthermore, $\sigma_{o,+}$ is governed by the impurity in the moderate-conductivity regime ($\sigma_{xx} \sim 10^{4-6} \, \Omega^{-1}\text{cm}^{-1}$) and by the host-lattice in the high-conductivity regime ($\sigma_{ZZ} \gtrsim 10^6 \, \Omega^{-1}\text{cm}^{-1}$). The same holds true for $\sigma_{e,-}$ and $\sigma_{o,-}$. Most FMs are in the moderate-conductivity regime at room temperature; therefore, tailoring impurity scattering is feasible for controlling $\boldsymbol{\sigma^{\perp}}$.

Finally, we discuss the role of spin-dependent scattering on $\boldsymbol{\sigma_{ij}^{\perp}}$. According to Ref. [22], perpendicularly polarized spin current can be transported by a spin-superimposed state with the same wavevector, namely, $a|\boldsymbol{k}\uparrow\rangle + b|\boldsymbol{k}\downarrow\rangle$. Although this role has been investigated for intrinsic mechanisms, it can be extended to extrinsic mechanisms. The superposition of a spin-conserve channel and a spin-flip channel transports the perpendicularly polarized spin current, as depicted in Fig. 3. For example, we consider skew-scattering with a spin-flip (Fig. 1(c)). This term is associated with the scattering T-matrix $T_{k\sigma;k'\sigma'} = J_{kk'}^{p}\delta_{\sigma',\sigma} + t_{kk'}^{\perp}\delta_{\sigma',\bar{\sigma}}$, where $t_{kk'}^{\perp}$ is the spin-flip term of $V^{sd}G^d V^{ds}$, which represents the scattering state $|\boldsymbol{k}\uparrow\rangle = (1 + TG^s)|\boldsymbol{k}'\uparrow\rangle = \sum_k\{(\delta_{k,k'} + J_{kk'}^{p}G_{k'\uparrow}^{s})|\boldsymbol{k}\uparrow\rangle + t_{kk'}^{\perp}G_{k'\uparrow}^{s}|\boldsymbol{k}\downarrow\rangle\}$. This yields $\sigma_{ij}^{\mu} \propto \sum_{k,k'} J_{kk'}^{p} t_{kk'}^{\perp} k_i k_j' \neq 0$. Note that the scattering state must have left-right asymmetry ($T_{k\sigma;k'\sigma'} \neq T_{-k\sigma;k'\sigma'}$) to avoid cancellation.

In conclusion, we have shown that spin-dependent scattering can induce perpendicularly polarized SHE with a reasonable order of magnitude, comparable with not only the host-lattice contribution but also the conventional (longitudinally polarized) SHEs. The spin-dependent scattering contribution can be dominant in common FMs in the moderately conductive region ( $10^{4-6}\,\Omega^{-1}\text{cm}^{-1}$ ). These results signify the importance of alloy engineering for controlling the SHE in FMs. To optimize the alloy constitution, an *ab initio* calculation using the coherent potential approximation is an area for future work.

We have also revealed the role of spin-dependent scattering, in which the superposition of a spin-conserve channel and the spin-flip channel is responsible for the perpendicularly polarized spin current. Although these spin currents may rapidly disappear in actual FMs, they exert a self-induced torque owing to the conservation law. The quantitative evaluation of this torque is critical and should be conducted in future studies.

**Acknowledgement**:

This work was supported by JST SPRING, Grant Number JPMJSP2114. Y.Y. acknowledges support from GP-Spin at Tohoku University.

Table I: List of coefficients $A_\sigma^{(x)}$ and $B_\sigma^{(x)}$ in Eqs. (2)–(5). The superscript denotes its microscopic mechanism shown in Fig. 1. The two rightmost columns are the estimation values in common ferromagnetic metals (FMs).

| (x) | $A_\sigma^{(x)}$ | $B_\sigma^{(x)}$ | $A_\sigma^{(x)}$ @ FM | $B_\sigma^{(x)}$ @ FM |
|---|---|---|---|---|
| (a) | – | $4\dfrac{\varepsilon_\sigma^{\rm F}}{\Delta_{\rm s}}\zeta_\sigma^{\rm F}$ | – | $4E_{\rm F}/\Delta_{\rm s}$ <br> $\sim 10^{1-2}$ |
| (b) | $\dfrac{5}{2}\zeta_\sigma^{\rm F}$ | $\dfrac{5}{4}(3+2\beta_\sigma^{\rm F})\zeta_\sigma^{\rm F}$ | 2.5 | 6.25 |
| (c) | $\dfrac{\varepsilon_\sigma^{\rm F}-\zeta_\sigma^{\rm F}\varepsilon_{\bar\sigma}^{\rm F}}{\varepsilon_\sigma^{\rm F}-\varepsilon_{\bar\sigma}^{\rm F}}\tan\delta_{1\sigma}$ | – | $-0.1$ | – |
| (d) | $(3+2\beta_\sigma^{\rm F})\zeta_\sigma^{\rm F}\tan\delta_{1\sigma}$ | – | $-0.5$ | – |

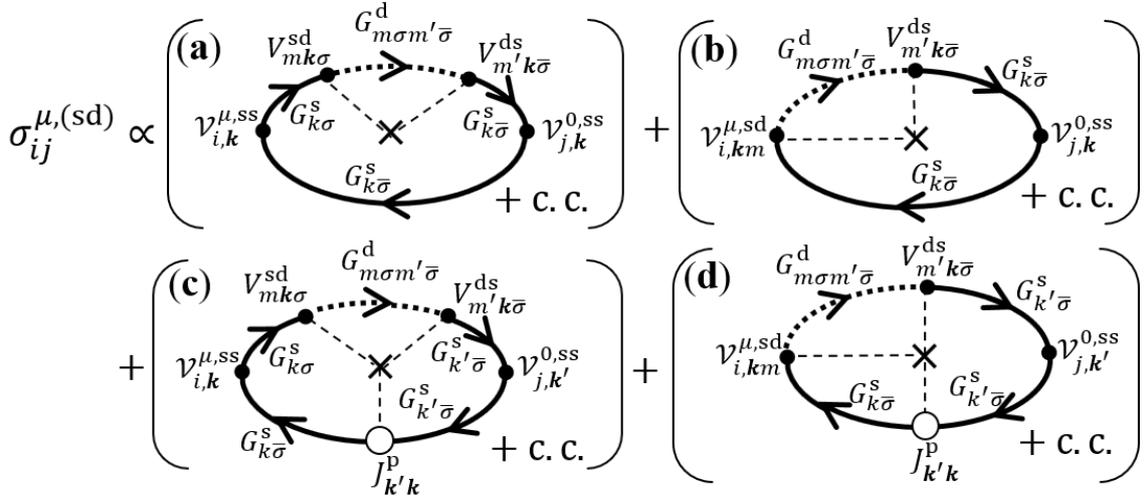

Fig. 1: (Color online) Diagrammatic expressions of $\sigma_{ij}^\mu$ ($\mu = X, Y$) under the averaged T-matrix approximation up to $O(n_{\rm imp})$ with the (a) coherent term, (b) anomalous velocity (AV) term, (c) current vertex correction (CVC) term, and (d) term with both CVC and AV. Here, $\bar\sigma = -\sigma$, and $m' = m + \sigma$.

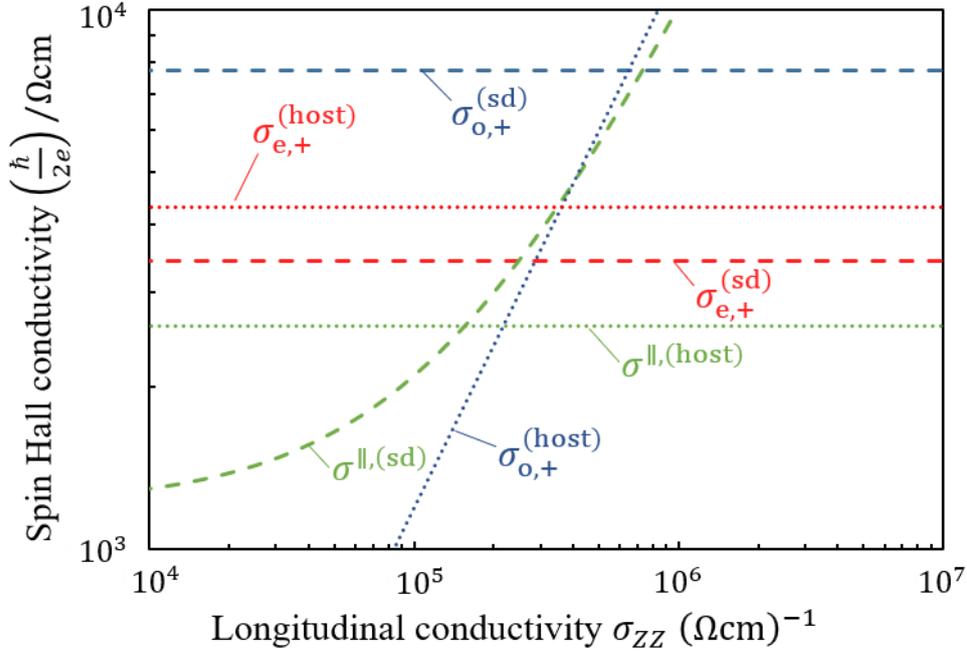

Fig. 2: (Color online) Perpendicularly polarized spin Hall conductivity as a function of the longitudinal conductivity. $\sigma^{(\mathrm{sd})}$ ($\sigma^{(\mathrm{host})}$) denotes the contribution from the spin-dependent scattering in the Fe-doped Ni (from the intrinsic and spin-independent scattering in the Ni host-lattice). The correspondence with the tensor components is given in Eq. (1). $\sigma^{\parallel}$ is the conventional (polarized parallel) spin Hall conductivity.

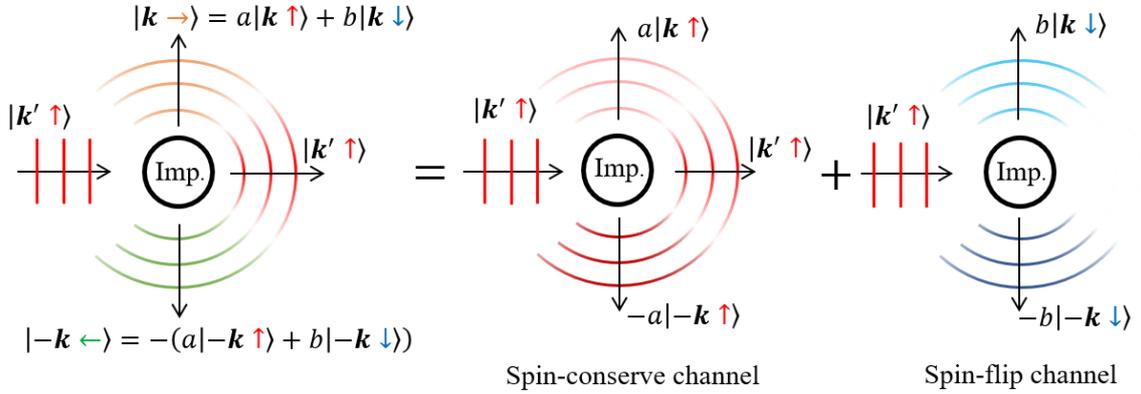

Fig. 3: (Color online) Schematics of the superposition of the spin-conserve and spin-flip channels, $(1 + TG)|\mathbf{k}'\uparrow(\downarrow)\rangle = \sum_{k}\{(\delta_{\mathbf{k}',\mathbf{k}} + a_{\mathbf{k}\mathbf{k}'})|\mathbf{k}\uparrow(\downarrow)\rangle + b_{\mathbf{k}\mathbf{k}'}|\mathbf{k}\downarrow(\uparrow)\rangle\}$. This superposition leads to a perpendicularly polarized spin current $\propto \langle \mathbf{k}\rightarrow|\mathbf{k}\hat{\sigma}^{\mu}|\mathbf{k}\leftarrow\rangle$. Here, we assume an asymmetric scattering such that $a_{-\mathbf{k}\mathbf{k}'} = -a_{\mathbf{k}\mathbf{k}'}$ ($b_{-\mathbf{k}\mathbf{k}'} = -b_{\mathbf{k}\mathbf{k}'}$).